\documentclass[fleqn,10pt]{wlscirep}
\usepackage[utf8]{inputenc}
\usepackage[T1]{fontenc}

\usepackage{booktabs}
\usepackage{tabularx}
\usepackage{array}
\usepackage{makecell}
\title{Emergent Generalization by Representation Learning in Artificial Neural Networks}

\author[1,2,*]{Hardik Rajpal}
\author[2]{Dan Goodman}

\affil[1]{I-X Centre for AI in Science, Imperial College London, W12 0BZ, UK}
\affil[2]{Department of Electrical and Electronic Engineering, Imperial College London, SW7 2AZ, UK}

\affil[*]{h.rajpal15@imperial.ac.uk}


\begin{abstract}
Dimensionality reduction has proven powerful for identifying neural manifolds, which are low-dimensional structures underlying high-dimensional neural activity. These low-dimensional representations have improved the interpretability of population-level coding. Yet whether such low-dimensional representations are biologically relevant and confer functional advantages in learning systems, or merely reflect neuron-level activity, remains contested in neuroscience. We show that an explicit information bottleneck forcing a recurrent neural network to learn a low-dimensional representation is necessary for rotational and out-of-distribution generalisation in a time-series prediction task. Using information-theoretic measures of causal emergence, we characterise the dynamics of this representation across the memorisation-to-generalisation transition, finding a non-monotonic trajectory which shows an initial decrease, a minimum, and a subsequent rise to a maximum, even as prediction loss falls monotonically. This trajectory scales with task complexity, and the magnitude of emergent structure reliably predicts generalisation performance. Analysis of hippocampal CA1 and medial prefrontal activity in rats learning an alternating maze task reveals analogous non-monotonic emergence dynamics that track behavioural performance. Together, these findings indicate that the ability of neural networks to learn compact, distributed and emergent representations confers a functional advantage for generalisation, supporting a causal role for learned representations in cognition.

\end{abstract}
\begin{document}

\flushbottom
\maketitle
\thispagestyle{empty}

\section*{Introduction}
With the advent of large-scale neural recordings, vast amounts of high-dimensional neural data recorded at high temporal resolution have become available from a variety of different brain regions and species. To interpret these complex datasets, dimensionality reduction techniques have been used to identify low-dimensional structures underlying high-dimensional neural activity~\cite{cunningham2014dimensionality}, which are referred to as neural manifolds~\cite{perich2025neural}. With repeatable experimental paradigms, stable and persistent neural manifolds have been identified across different brain regions corresponding to different cognitive and behavioural tasks, such as olfaction~\cite{stopfer2003intensity,mazor2005transient}, motor control~\cite{golub2018learning,churchland2012neural}, spatial navigation~\cite{kim2017ring}, and even intrinsic neural dynamics~\cite{chaudhuri2019intrinsic}. The presence of these low-dimensional manifolds across tasks and species has led to greater interpretability of neural dynamics and has unveiled the geometric structure of neural computations.

More recently, experiments have sought to establish a direct causal relationship between neural manifolds and behaviour by perturbing neurons within them and observing the effects on behaviour. For example, head orientation of the fruit fly can be manipulated optogenetically by perturbing the neurons in the ring attractor manifold~\cite{kim2019generation}. Similar perturbation experiments have been performed to impact complex behaviours such as motor timing~\cite{yang2026integrator} and adaptive task learning~\cite{cho2023long} in mice. Complementarily, studies using fixed linear decoders have shown that behavioural variables can be reliably decoded from low-dimensional manifolds in the same animal up to years after initial training in monkeys~\cite{gallego2020long} and across twin \textit{C. elegans}~\cite{brennan2019quantitative}. These studies suggest that the low-dimensional manifolds may not merely be descriptive epiphenomena of neural activity, but rather have quantifiable causal power in driving behaviour.

However, there still exists a strong and consistent debate about the methods, interpretation, and the necessity of the neural manifold view in neuroscience. On the methodological side, there are concerns about the horseshoe effect in dimensionality reduction techniques, which can lead to spurious low-dimensional structure in the data~\cite{proix2022misinterpreting}. This has led to the development of better population controls~\cite{elsayed2017structure} and null models~\cite{humphries2021spectral} to test the significance of the low-dimensional structure in the data. A separate line of methodological criticism has been raised about the interpretability of linear dimensionality reduction techniques, which may not be able to capture the widely prevalent non-linear structure in the data~\cite{de2023common}. Finally, recent studies have explored how a small group of non-representative neurons can skew the representation without affecting the decoding accuracy, which questions the positioning of neural manifolds as a representative summary of the neural population activity~\cite{bertram2026decoding}. This debate is continually evolving, and new methods and techniques are being developed to address these concerns, which will further help to clarify the role of neural manifolds in neuroscience.

In our study, we approach the question of the relevance of neural manifolds from a functional perspective by asking if the ability to learn low-dimensional representations of high-dimensional inputs confers any functional advantages in learning systems. Building on the existing literature on the information bottleneck in deep learning~\cite{tishby2000information}, we explore the relationship between the learning of low-dimensional representations and the ability to generalise to out-of-distribution test data. It has been argued that robust generalisation requires compressing away task-irrelevant information while preserving task-relevant structure~\cite{tishby2000information,shwartz2017opening}. This view has produced various advances in the field of task generalisation in deep learning, most notably in uncovering the relationship between compression and invariance~\cite{achille2018emergence}, variational inference~\cite{alemi2016deep}, the necessity of invariant risk minimization~\cite{arjovsky2019invariant} and contrastive learning methods to avoid shortcut learning~\cite{li2025contrastive}. 

However, there have been three main criticisms of the relationship between compression and generalisation. First, it has been shown that compression is not necessary for generalisation, and that networks can generalise without compressing the input data~\cite{saxe2019information}. The second criticism is that the exponential dependence of sample complexity on the inverse of target error makes it impractical for practical applications~\cite{hafez2019compressed}. Finally, it has been shown that the relationship between compression and generalisation is not monotonic, and the networks transition between a rapid fitting phase (memorisation) and a slow compression phase (generalisation)~\cite{goldfeld2018estimating}. This phenomenon has been referred to as grokking, and has been observed in various settings, including image classification~\cite{humayun2024deep} and algorithmic tasks~\cite{power2022grokking,liu2022towards}. However, over-compression in the generalisation phase can enhance discriminability at the expense of transferability~\cite{cui2022discriminability}. While we do not address the first two criticisms in this work, we do explore the non-monotonic relationship between compression and generalisation further in the context of next-step prediction and identify that the grokking transition is associated with an increase in the emergent structure of the learned latent representation, which in turn is correlated with out-of-distribution performance.

In this work, we combine the insights from the literature on neural manifolds and the information bottleneck to address the functional need of extracting low-dimensional representations for any learning system embedded in a high-dimensional world. This question is relevant to neuroscience since the sensory signals that the brain receives are high-dimensional, while it manages to perform contextually similar tasks in different environments. This ability to generalise across different contexts is a key feature of intelligence, and has been studied in the context of meta-learning~\cite{finn2017model} and abstraction~\cite{lake2017building}. Separately, in the field of statistical physics and dynamical systems, low-dimensional representations correspond to coarse-grainings of the system. In certain cases, these coarse-grainings can be shown to be causally emergent, such that the representation has greater information about its own future state than the sum of the marginal predictive information provided by each part separately~\cite{hoel2013quantifying,flack2017coarse,rosas2020reconciling}. This suggests that when the optimal low-dimensional representations learned by the network are causally emergent, they may be able to leverage the enhanced predictive power of the low-dimensional representation to generalise to out-of-distribution test data. In this work, we test whether the causal emergence of learned latent representations predicts such out-of-distribution generalisation. We argue that this ability to generalise to out-of-distribution data by extracting low-dimensional abstract representations may explain the functional need for neural manifolds in biological neural networks. 

To establish this relationship between neural manifolds and task generalisation, we use a simple reservoir computing architecture with a bottleneck layer to predict the next time step of a high-dimensional time series generated from a low-dimensional latent dynamical system. We show that by learning an overcomplete low-dimensional representation at the bottleneck layer, the network can generalise both to random orthogonal projections of the trained latent dynamical systems as well as to held-out dynamical systems that are out of distribution from the training data. By characterising the information-theoretic properties of the learned latent representation, we show that the learned latent representation becomes causally emergent during training, and explain the network's ability to generalise to out-of-distribution test data. Finally, we show that a similar non-monotonic behaviour of the causal emergence of latent representations is observed in CA1 and medial PFC of rats during a spatial learning task, which suggests that the learning of emergent neural manifolds may be a key feature of learning in biological neural networks.

\section*{Results}
\begin{figure}[!h]
\centering
\includegraphics[width=0.8\linewidth]{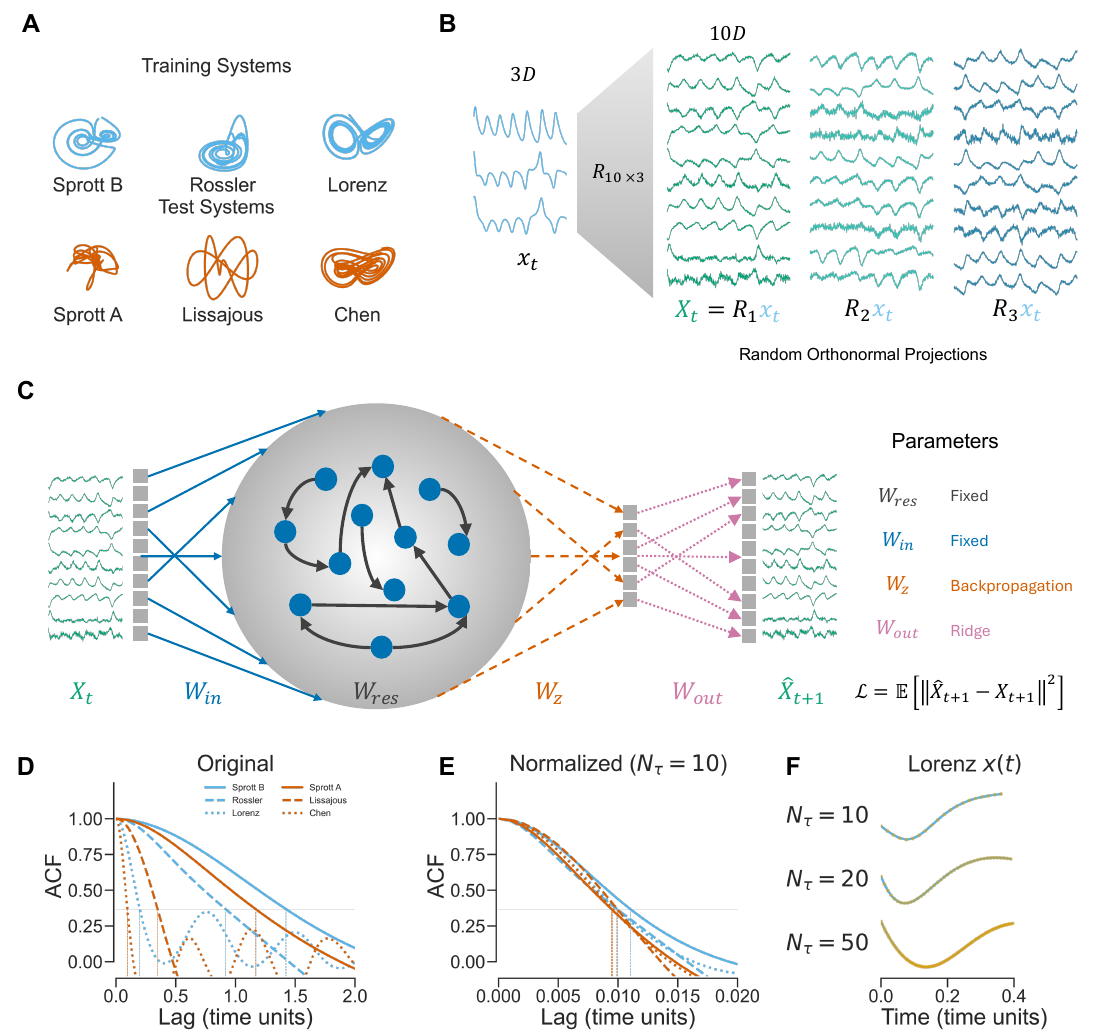}
\caption{\textbf{A} The six dynamical systems used in the study. Data generated using the training systems was used to train the model. Test systems are used to evaluate the OOD generalisation performance of the model. \textbf{B} The input data is generated from a 3-dimensional latent dynamical system, which is then projected to a 10-dimensional space using a random orthogonal projection. The network is trained on data generated from a subset of the different dynamical systems with different orthonormal projections. \textbf{C} The input is projected to the reservoir network, which is a fixed recurrent network. The states of the reservoir network are then projected to a low-dimensional bottleneck layer, which is then used to predict the next time step of the input data using a ridge linear readout layer. The bottleneck layer is trained using backpropagation to learn a low-dimensional representation that minimises the MSE after the ridge regression. \textbf{D} The different dynamical systems have different autocorrelation times, which vary the complexity of the task. \textbf{E} We normalise the timescales of the different dynamical systems by subsampling and keeping $N_{\tau}$ timesteps of the generated data before the autocorrelation of the data drops below $1/e$. \textbf{F} Smaller values of $N_{\tau}$ result in more difficult tasks, by reducing the amount of information available for next step prediction.}
\label{fig:setup}
\end{figure}

In this work, we focus on the time series prediction task in a high-dimensional setting, where the input data is a high-dimensional time series $X_t$ generated from a low-dimensional latent dynamical system $x_t$. We generate training and testing data from 6 different low-dimensional dynamical systems (see Figure~\ref{fig:setup} A), which are then projected to a high-dimensional space using a random orthogonal projection $R$, with some added noise (see Figure~\ref{fig:setup} B). The network is trained on data generated from the training systems with different orthonormal projections, and then tested on the held-out test systems, which are out of distribution from the training data (see Figure~\ref{fig:setup} A). The network is trained to predict the next time step of the input time series. For the purposes of this study, we use a modified Reservoir Computing Architecture (see Figure~\ref{fig:setup} C), with a bottleneck layer which is trained using backpropagation. While the linear readout layer is trained using ridge regression to predict the next time step of the input data, the bottleneck layer is trained to learn a low-dimensional predictive latent representation of the input data. This architecture provides a simple and interpretable framework to study the functional advantages of representation learning in neural networks.

The different dynamical systems have inherently different timescales, which modulates the complexity of the task (see Figure~\ref{fig:setup} D). For instance, the Lorenz attractor shows a faster decay of autocorrelation as compared to the Rössler attractor, which makes it more difficult to predict the next time step of the input data. To equalise this difficulty across systems, we subsample each time series so that exactly $N_{\tau}$ samples remain before the autocorrelation drops below $1/e$ (see Figures~\ref{fig:setup} D,E). This allows us both to modulate the complexity of the task, as well as ensure that the different dynamical systems are equally difficult to predict. For smaller values of $N_{\tau}$, the task becomes more difficult (see Figure~\ref{fig:setup} F), as the network has to learn to predict the next time step of the input data with less information about the past history of the input data. 

\subsection*{Information Bottleneck enables Rotation Invariant OOD Generalisation}
\begin{figure}[!h]
    \centering
    \includegraphics[width=0.8\linewidth]{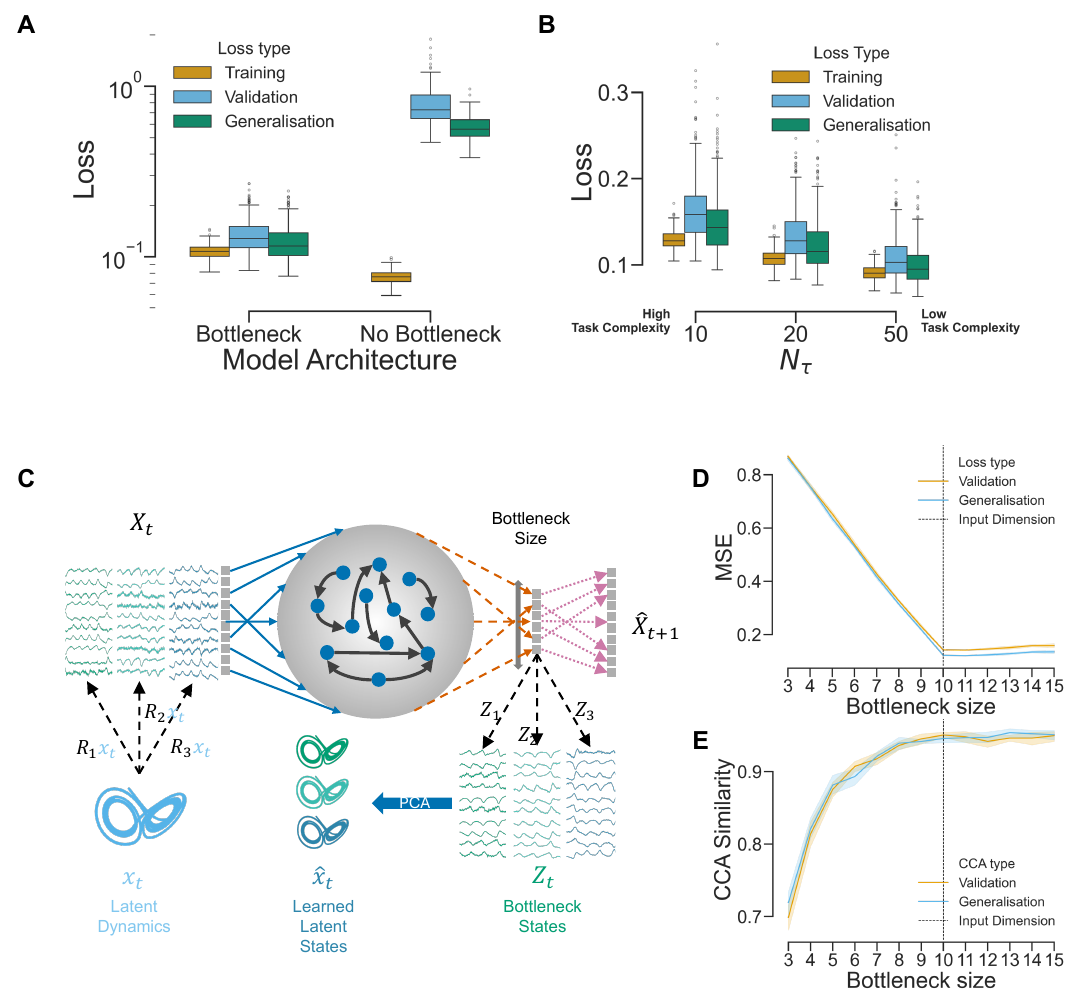}
    \caption{\textbf{A} Comparing the next step prediction loss of the network with and without the bottleneck layer, for a fixed task complexity ($N_{\tau} = 20$). The network without the bottleneck layer minimises training error but is not able to generalise to random rotations of the training attractors or to the held-out dynamical systems. \textbf{B} Task complexity modulates the training, validation and generalisation loss. Prediction error increases with increasing task complexity, which is controlled by subsampling the input data. \textbf{C} Schematic representation of the overcomplete latent representation learned by the bottleneck layer. The three-dimensional PCA projection of the bottleneck states shows the recovered latent states of the dynamical system, which are rotation invariant. \textbf{D} Next step prediction loss for unseen rotations of training attractors and held out dynamical systems as a function of the size of the bottleneck layer. The loss decreases with increasing size of the bottleneck layer, and reaches a minimum when the size of the bottleneck layer matches the dimensionality of the input. \textbf{E} Similarity between the low-dimensional representation learned by the bottleneck layer and the true latent states of the dynamical system using Canonical Correlation Analysis (CCA). The similarity increases with increasing size of the bottleneck layer, and reaches a maximum when the size of the bottleneck layer matches the dimensionality of the input.} 
    \label{fig2:rotations}
\end{figure}

First, we show that the recurrent reservoir network with a bottleneck layer generalises both to unseen rotations of the training attractors and to held-out attractors that were never seen during training. Without a bottleneck layer, the network is able to learn the training data, but is not able to generalise to random rotations of the training attractors or to the held out dynamical systems (see Figure~\ref{fig2:rotations} A). This suggests that the bottleneck layer and representation learning are a necessary feature for enabling generalisation. The network's prediction accuracy decreases with increasing task complexity using subsampling $N_{\tau}$ points before the autocorrelation decay. We find that all training, validation and generalisation loss increase with increasing task complexity (see Figure~\ref{fig2:rotations} B).

However, the size of the bottleneck layer is a key parameter that controls the network's ability to generalise to random rotations of the latent dynamical systems in the training set. The main challenge in this task is not just the extraction of the low-dimensional latent structure of the input data, but also to predict any random projection of the latent states with a fixed readout layer. This requires the network to learn a rotation-invariant representation of the latent states, which is key for out-of-distribution generalisation (see Figure~\ref{fig2:rotations} C).

As we increase the size of the bottleneck layer, the next step prediction loss monotonically decreases for both unseen rotations of training attractors as well as for the held out dynamical systems, and reaches a minimum when the size of the bottleneck layer matches the dimensionality of the input (see Figure~\ref{fig2:rotations} D). Any further increase in the size of the bottleneck layer causes a slight increase in the next step prediction loss. The true dimensionality of the bottleneck layer is still low rank and matches the dimensionality of the latent dynamical system, which suggests that the bottleneck learns an overcomplete low-dimensional basis of the dynamics, which allows it to learn a general solution in a common latent space. We can extract the learned low-dimensional representation by performing a PCA and projecting the bottleneck states to a three-dimensional space.

Using Canonical Correlation Analysis (CCA), we then measure how closely the learned low-dimensional representation matches the true latent states of each dynamical system. CCA similarity grows with bottleneck size and peaks when the bottleneck is as wide as the input (see Figure~\ref{fig2:rotations} E). This confirms that the bottleneck layer is able to learn the true latent states of the dynamical systems, both for random rotations of the training attractors as well as for the held-out dynamical systems.

\subsection*{Generalisation Loss exclusively depends on Task Complexity}
\begin{figure}[!h]
    \centering
    \includegraphics[width=\linewidth]{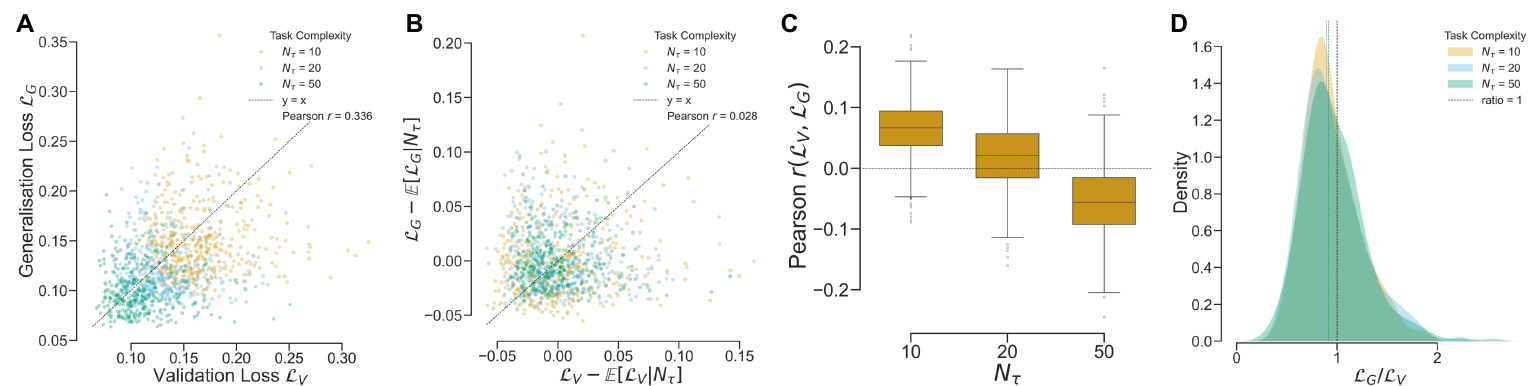}
    \caption{\textbf{A} The prediction loss for unseen rotations of the training attractors (validation loss) and the prediction loss for held out dynamical systems (generalisation loss) are weakly correlated (Pearson r $=0.336$) across different trained models for different task complexities. \textbf{B} When conditioning on the task complexity, the correlation drops to $0.028$, which suggests that task complexity modulates the correlation between $\mathcal{L}_{V}$ and $\mathcal{L}_{G}$. \textbf{C} The box plots show the Pearson correlation between $\mathcal{L}_{V}$ and $\mathcal{L}_{G}$ across different trained models for different task complexities. The correlation is low and positive for high task complexities, and low and negative for low task complexities. \textbf{D} The histogram of the ratio of generalisation loss to validation loss across different trained models and task complexities. The median ratio is slightly smaller but close to 1, which indicates robust generalisation capacity of the network across task complexities.}
    \label{fig3:loss_robustness}
\end{figure}

Next, we show that the network's ability to extract the low-dimensional structure depends only on the task complexity and the chaoticity of the latent dynamical system being predicted. This decouples the validation loss for unseen rotations of the training attractors from the generalisation loss for held out dynamical systems. The prediction loss for unseen rotations of the training attractors (validation loss) and the prediction loss for held out dynamical systems (generalisation loss) are not correlated, once the task complexity is conditioned on (see Figure~\ref{fig3:loss_robustness} A,B). When comparing the Pearson correlation between validation loss $\mathcal{L}_{V}$ and generalisation loss $\mathcal{L}_{G}$ across different trained models for different task complexities, we find that the correlation is low and positive for high task complexities, and low and negative for low task complexities (see Figure~\ref{fig3:loss_robustness} C). Furthermore, when looking at the ratio of generalisation loss to validation loss, across different trained models and task complexities, we find the ratio is consistently close to 1, which suggests that the generalisation ability of the network is robust and not dependent on learning a specific attractor (see Figure~\ref{fig3:loss_robustness} D).

\subsection*{Learned Latent Representations are Causally Emergent}
\begin{figure}[!h]
    \centering
    \includegraphics[width=0.8\linewidth]{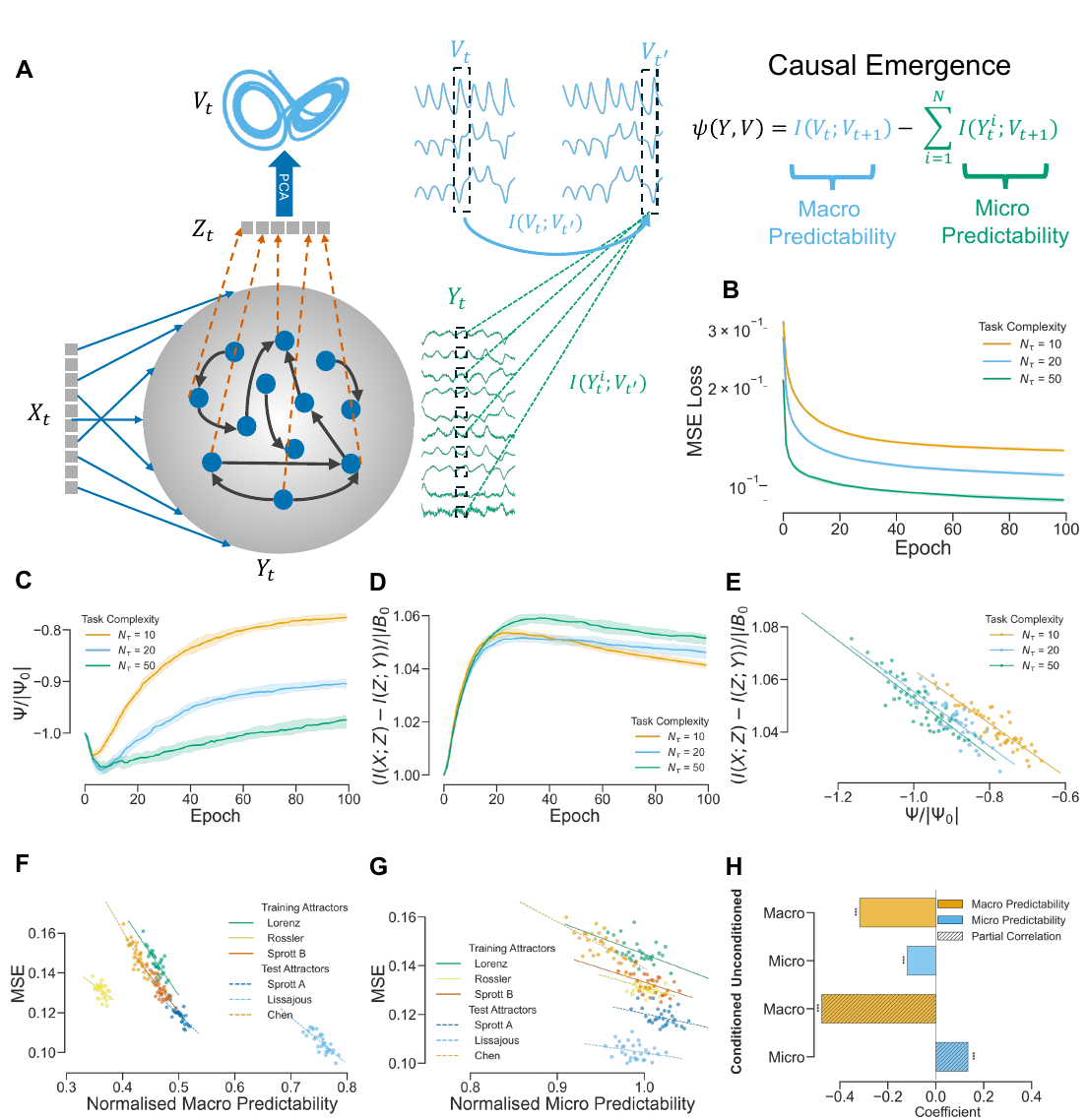}
    \caption{\textbf{A} Schematic representation of the micro states $Y_t$ of the reservoir network and the macro states $V_t$, which are the three-dimensional PCA projection of the bottleneck states. The $\Psi$ measure quantifies the degree to which the macro states are more predictive of their own future state than the sum of the information that each micro state carries about that future state. \textbf{B} The MSE decreases monotonically during training across the epochs, while the minimum MSE is controlled by the task complexity $N_{\tau}$. \textbf{C} The $\Psi$ measure of causal emergence initially decreases during the early stages of learning and then, after reaching a minimum, increases to its maximum value during the generalisation phase. A higher value of $\Psi$ is observed for more complex tasks. \textbf{D} A similar non-monotonic behaviour is observed for the Information Bottleneck loss. The IB loss initially increases during the early stages of learning and then starts decreasing during the generalisation phase. \textbf{E} The normalised $\Psi$ measure post-training is negatively correlated with the IB loss. \textbf{F} The normalised macro predictability is negatively correlated with the prediction loss, and its magnitude varies across the different attractors. \textbf{G} The normalised micro predictability is also negatively correlated with the prediction loss. \textbf{H} While both components show negative correlation with the prediction loss (across both training and held out attractors), the combined linear model shows that the effect goes in opposite directions when the two components are considered together.}
    \label{fig4:emergence}
\end{figure}

Using information-theoretic measures of causal emergence, we show that the learned latent representations in the bottleneck layer become increasingly causally emergent during learning, and that their degree of emergence is correlated with generalisation performance. The $\Psi$ measure of causal emergence quantifies the degree to which any coarse-grained representation of a system is predictive of its own future state (macro predictability), as compared to the sum of the information that each individual part of the system carries about that future state (micro predictability) (see Equation~\ref{eq:Psi}). This difference between macro and micro predictability quantifies the degree to which the coarse-grained representation is more than the sum of its parts, and thus emergent. A simple example of causal emergence is the flocking behaviour, where the future direction of the flock is better predicted by the centroid of the flock, than by the individual birds in the flock~\cite{rosas2020reconciling}. In neuroscience, this behaviour has been observed~\cite{brunel2008sparsely} and quantified~\cite{rajpal2026quantifying} in sparsely synchronised neuronal oscillations, where the future state of the oscillation is better predicted by the phase of the oscillation, than by the irregularly firing individual neurons in the oscillation. In this study, we use the three-dimensional PCA projection of the bottleneck states (learned latent representation) as the coarse-grained representation of the system, and the neurons of the reservoir as the parts of the system (see Figure~\ref{fig4:emergence} A). 

\begin{equation}
    \Psi = \underbrace{I(V_t; V_{t+1})}_{\text{Macro predictability}} - \underbrace{\sum_{i=1}^{N} I(Y^i_t; V_{t+1})}_{\text{Micro predictability}}
    \label{eq:Psi}
\end{equation}

We estimate $\Psi$ at each epoch of training and track how it changes during learning, while the MSE monotonically decreases (see Figure~\ref{fig4:emergence} B). After normalising the measure by the absolute value at the first epoch of training, we find a non-monotonic behaviour of the measure during learning, where the emergent behaviour of the latent representation initially decreases during the early stages of learning (memorisation phase) and then increases after reaching a critical point in learning (generalisation phase), which coincides with the grokking phenomenon described in recent work~\cite{power2022grokking,clauw2024information} (see Figure~\ref{fig4:emergence} C). This suggests that the network switches to learning more emergent latent representations during the grokking transition. Furthermore, we find that the degree of emergent behaviour exhibited by the trained network increases with the complexity of the task (see Figure~\ref{fig4:emergence} C), indicating that more difficult prediction tasks require the learning of more emergent latent representations during training.

We compare the non-monotonic behaviour of the measure of causal emergence with the Information Bottleneck (IB) measure~\cite{tishby2000information}, which compares the amount of predictive information preserved in the latent representation about the target output while compressing the irrelevant information about the input. We observe a similar non-monotonic behaviour of the IB loss where it initially increases during early stages of learning and then decreases during the generalisation phase (see Figure~\ref{fig4:emergence} D). However, the $\Psi$ transition precedes the IB transition during training. Across the trained models, we find that the normalised $\Psi$ measure post-training is negatively correlated with the IB loss, which suggests that generalisation is associated with the learning of both compressed and emergent latent representations (see Figure~\ref{fig4:emergence} E).

Finally, to test how the degree of emergence relates to generalisation performance, we decompose the $\Psi$ measure into its macro and micro predictability components and relate them to the next step prediction loss. We find that these measures vary significantly across different attractors, as the predictability is bound by the inherent chaoticity of the dynamical system (see Figures~\ref{fig4:emergence} F,G). Therefore, we first normalise the macro and micro predictability by that of the true latent attractor. We then use a linear mixed effects model to account for the random effect of the attractor on the prediction loss. We find that the normalised macro predictability component of the $\Psi$ measure is negatively correlated with the prediction loss, while a significant positive correlation between the micro predictability component and the prediction loss is observed only after conditioning on the macro component (see Figure~\ref{fig4:emergence} H). While high macro predictability is associated with stability of the representation, low micro predictability is associated with highly distributed representations, where no single component can predict the behaviour of the macro. These results indicate that the networks that learn both maximally predictive as well as highly distributed latent representations are able to generalise to out-of-distribution test data. The $\Psi$ measure summarises both these effects and provides an effective proxy for the optimality of the learned latent representation for generalisation.

\subsection*{Emergent Representations in spatial learning tasks in rats}
\begin{figure}[!h]
    \centering
    
    \includegraphics[width=0.7\linewidth]{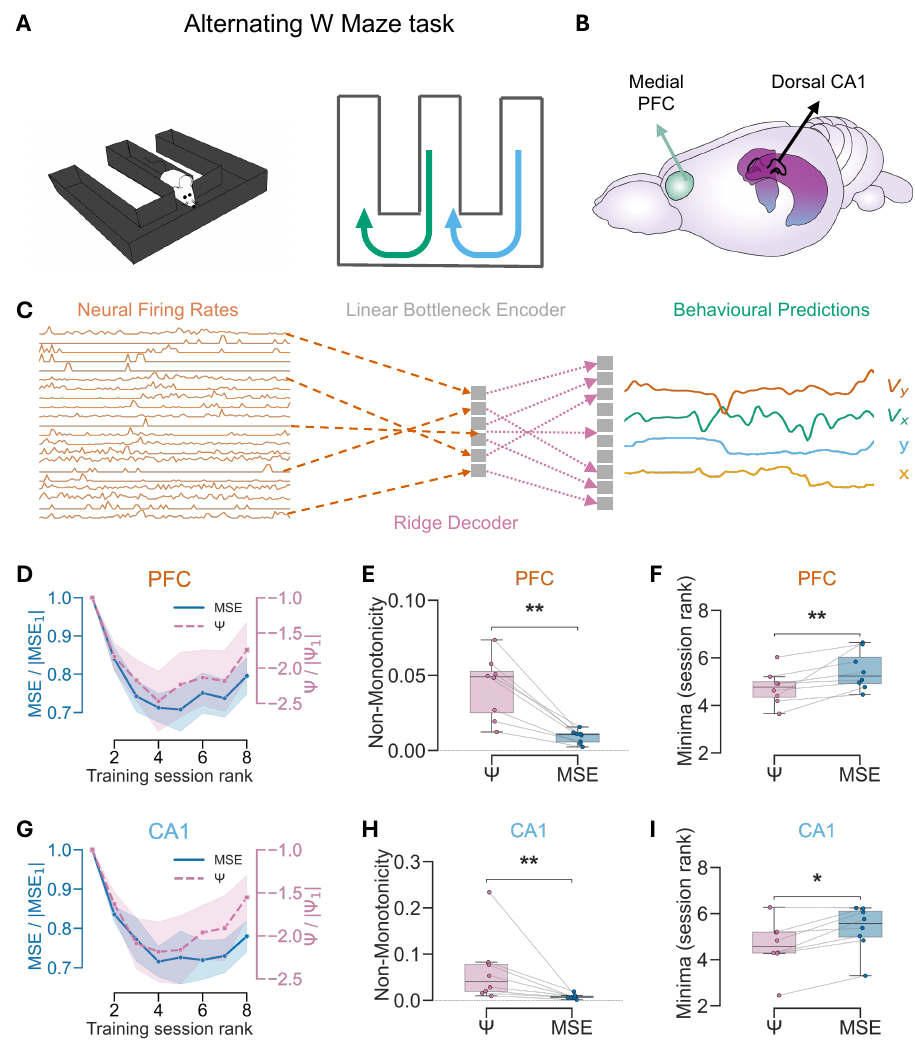}
    \caption{\textbf{A} Schematic representation of the W maze spatial learning task. The rats are trained to learn a continuous alternation task in a W maze. \textbf{B} The sagittal view of the rat brain showing the location of the CA1 and medial PFC regions where the neural activity is recorded during the task. \textbf{C} The setup for representation learning in the neural data. The preprocessed firing rates are passed through a simple linear bottleneck encoder and ridge readout decoder to predict the behavioural measures (position and velocity) of the rats in the next timestep. \textbf{D,G} The $\Psi$ measure of causal emergence and the decoder MSE during learning in PFC (\textbf{D}) and CA1 (\textbf{G}). Both measures are normalised by their respective absolute values in the first session. \textbf{E,H} Non-monotonicity, as measured by the quadratic coefficient of the linear model fit to the $\Psi$ and MSE measures during learning in PFC (\textbf{E}) and CA1 (\textbf{H}). The quadratic coefficient is significantly larger for the $\Psi$ measure as compared to the MSE in both regions. \textbf{F,I} Using the fitted parameters of the linear model, the estimated minima of the $\Psi$ measure significantly precede the minima of the MSE in both PFC (\textbf{F}) and CA1 (\textbf{I}).}
    \label{fig5:emergence_mice}
\end{figure}

To test the validity of our findings beyond synthetic data and artificial neural networks, we test whether the non-monotonic behaviour of the measure of causal emergence during learning is also observed in biological neural networks. We analyse the data from a recent study which recorded the activity of neurons from the dorsal CA1 and the medial PFC of rats during a W-maze spatial learning task~\cite{shin2019dynamics}. The dataset is obtained from 8 rats, which are trained to learn a continuous alternation task in a W maze, where the rats are rewarded for sequentially visiting the middle and the top arm when starting from the bottom arm of the maze (see Figure~\ref{fig5:emergence_mice} A). The rats are trained for over 8 training sessions, and the behavioural measures (position and velocity) and performance measures (correct and incorrect trials) are recorded along with the neural activity from the CA1 and the medial PFC (see Figure~\ref{fig5:emergence_mice} B). We use the preprocessed firing rates of the neurons from one of the regions (CA1 or medial PFC) as input to a simple network with a linear bottleneck encoder and ridge readout decoder, and train the network to predict the behavioural measures (position and velocity) of the rats in the next timestep (see Figure~\ref{fig5:emergence_mice} C). We then estimate the measure of causal emergence of the bottleneck states in the trained model for each of the 8 training sessions, and track how it changes along with the decoder prediction MSE during learning across sessions. 

We find that the measure of causal emergence exhibits a similar non-monotonic behaviour during learning, where it initially decreases during the early stages of learning and then increases in the later stages, which is consistent with our findings in the reservoir computing setup. This increase in the emergence measures precedes the increase in the decoder MSE in the final sessions, both in CA1 and the medial PFC (see Figures~\ref{fig5:emergence_mice} D,G). To confirm the difference in the non-monotonic behaviour between the emergence and the MSE, we fit a quadratic linear model to the measures for each subject and each brain region. We find that the quadratic coefficient is significantly larger for the emergence measure as compared to the MSE in both CA1 and the medial PFC, which suggests that the emergence measure shows a significant non-monotonic increase as compared to the MSE (see Figures~\ref{fig5:emergence_mice} E,H). Consequently, we also observe the minima of the emergence measure to significantly precede the minima of the MSE in both regions (see Figures~\ref{fig5:emergence_mice} F,I).

These results are in line with the previous findings that show a similar non-monotonic behaviour of the neural-latent correlation of the neural manifold during learning in the same dataset~\cite{wakhloo2026neural}. These results suggest that the learning of emergent latent representations may be a key feature of learning in biological neural networks. However, further work is needed to establish the link from emergence to generalisation in biological neural networks in carefully designed experiments.

\section*{Discussion}
In this study, we show that the ability to learn low-dimensional representations of high-dimensional inputs confers functional advantages in learning systems. Here, a simple reservoir computing architecture with a bottleneck layer is able to learn low-dimensional latent representations of high-dimensional input data, which allows it to predict both unseen projections of the training systems and entirely new dynamical systems. This ability was crucially dependent on the presence of the bottleneck layer, without which the network overfits to training data and is unable to generalise. In essence, we implement a meta-learning framework, where the network learns to extract the low-dimensional latent structure that generated the input data and thereby generalises to unseen data. Notably, the model shows zero-shot generalisation to held-out dynamical systems, indicating that the learned latent space is shared across systems rather than specific to the training attractors. The performance is robust across different dynamical systems and is modulated only by task complexity and the inherent chaoticity of the dynamical system being predicted.

This ability to generalise using a low-dimensional latent representation is associated with the enhanced predictability of the latent dynamics, which characterises the learned latent representation as causally emergent. We show that the degree of causal emergence of the learned latent representation is correlated with the network's ability to generalise to out-of-distribution test data. These results extend the previous work on bi-directional link between emergence and prediction~\cite{tolle2026evolving} to representation learning and OOD generalisation. Furthermore, we show that the measure of causal emergence exhibits a non-monotonic behaviour during the grokking transition described in recent work~\cite{power2022grokking,clauw2024information}. Finally, recordings from CA1 and medial PFC of rats learning a spatial alternation task show the same non-monotonic trajectory of emergence, with its minimum preceding that of the decoder MSE, suggesting that the formation of emergent neural manifolds may be a common feature of learning in artificial and biological networks.

While this study identifies a functional advantage of learning low-dimensional representations, it does not address the mechanistic question of how these representations are learned. In this study, we use a simple reservoir computing architecture with fixed weights and a bottleneck layer trained using backpropagation to learn low-dimensional latent representations. To investigate the mechanistic basis of representation learning, we need to explore the various plasticity mechanisms and biological constraints that may enable the learning of emergent neural manifolds in the brain.
Furthermore, while we show that the learned latent representations are causally emergent, it is unclear if this is a necessary condition for generalisation. Future work should explore whether such out-of-distribution generalisation can be achieved without learning a compressed and emergent latent representation.

Finally, although the model achieves zero-shot generalisation to held-out dynamical systems, it is unable to perform closed-loop prediction as a generative model. Further work is needed on how the latent representations can be refined online to enable generative predictions. Recent work has shown that intermediate refinement of latent representations can lead to optimal generalisation~\cite{fountas2026brain}. This study also opens new questions on the relationship of neural manifolds to causal emergence in the temporal domain. While geometric characterisations of neural manifolds have been extensively studied~\cite{chung2021neural}, the temporal dynamics of neural manifolds and their relationship to causal emergence remain an open question. In future work, we plan on exploring this relationship in more detail and investigating the mechanistic basis of learning emergent neural manifolds in biological neural networks.

\section*{Methods}
In this study, we use several dynamical systems to generate the input data, a modified reservoir computing architecture, and information-theoretic measures to quantify the emergence of learned latent representations. Here we describe in detail the methods used in this study. 

\subsection*{Dynamical systems and data generation}
We simulated 6 different three-dimensional dynamical systems ($x_t$) using RK4 integration. These attractors were further divided into three training attractors and three held-out test attractors. The training attractors used in this study were the chaotic Lorenz attractor, the Rössler attractor, and the Sprott B attractor. The held out test attractors were the chaotic Chen attractor, the Sprott A attractor, and the periodic Lissajous attractor. These attractors were chosen as they have different topological and dynamical properties, such as the autocorrelation time, Lyapunov exponent, and fractal dimension. The attractors were simulated for over 100,000 timesteps with an integration timestep of 0.001, and the first 1000 timesteps were discarded to remove any transients. See Appendix~\ref{appx:attractors} for the equations and parameters of the different attractors used in this study. The dynamics of the attractors are defined by the ordinary differential equation of a three-dimensional system $x(t)$,

\begin{equation}
    \frac{d x(t)}{dt} = f(x(t))
\end{equation}

In order to systematically modulate the complexity of the task, we downsampled the simulated attractor (integration timestep = 0.001) timeseries to keep exactly $N_{\tau}$ equally spaced timesteps of the attractor timeseries before the autocorrelation of the timeseries drops below $1/e$. This allows us to control the amount of information available to the network for predicting the next timestep of the input data, and thereby modulate the complexity of the task. The subsampled attractor timeseries were then projected to a 10-dimensional space using a random orthogonal projection $R$, with some added Gaussian noise $\epsilon$. The high-dimensional input data was then used to train and test the network.

\begin{equation}
    \begin{split}
    X_t &= R x_t + \epsilon \\
    \epsilon &\sim \mathcal{N}(0, \sigma^2 I)
    \end{split}
\end{equation}

\subsection*{Model architecture and training}
The model architecture used in this study is a modified reservoir computing architecture, which consists of a fixed input and recurrent reservoir layer, followed by a trainable bottleneck layer and a linear readout layer. The input data $X_t$ is projected to the reservoir layer using a fixed random projection $W_{in}$, and the states of the reservoir layer $Y_t$ are then projected to the bottleneck layer, $Z_t = W_z Y_t$, using a trainable linear transformation $W_z$. The bottleneck layer is then used to predict the next timestep of the input data using a linear readout layer $W_{out}$. The bottleneck layer is trained using backpropagation, while the linear readout layer is trained using ridge regression to predict the next timestep of the input data. The model is trained using mean squared error of the next step prediction loss for 100 epochs, using the Adam optimiser with a learning rate of 0.001 in PyTorch. This setup trains the bottleneck layer to learn a low dimensional representation optimal for next step prediction. The size of the bottleneck layer is varied to study its effect on generalisation. We use a batch size of 12, and each batch consists of 10,000 timesteps of a random rotation of the training attractors starting from different initial conditions. We use a washout length of 1000 timesteps. The validation and test sets are generated independently from the training set, and consist of random rotations of the training attractors and held-out test attractors, respectively.

The reservoir dynamics are defined by the following recurrent dynamics,
\begin{equation}
    \begin{split}
    Y_t &= h\tanh(W_{res} Y_{t-1} + W_{in} X_t + b) + (1-h)Y_{t-1}
    \end{split}
\end{equation}

Here, $h$ is the leak rate of the reservoir, which controls the timescale of the reservoir dynamics, and $b$ is the input bias. The reservoir weights $W_{res}$ are initialised from a normal distribution. We use the reservoir size of 500 across all experiments, and the hyperparameters of the reservoir, including the leak rate, spectral radius, sparsity, the input scaling and the input bias, are optimised using the Tree-structured Parzen Estimator (TPE) algorithm implemented in Optuna. The hyperparameters of the reservoir are optimised to minimise the validation loss on unseen rotations of the training attractors, and the best hyperparameters are used for all experiments.

\subsection*{Information theoretic measures}
We use the information-theoretic measures of causal emergence and information bottleneck to characterise the learned latent representations in the bottleneck layer. Both of these measures are based on mutual information, which quantifies the amount of information shared between two random variables. Here, we assume the states to be multivariate Gaussian random variables, thus capturing only linear dependencies. The mutual information between two multivariate Gaussian random variables $A$ and $B$ is defined as,

\begin{equation}
    I(A;B) = \frac{1}{2} \log \left( \frac{| \Sigma_A | | \Sigma_B |}{| \Sigma_{AB} |} \right)
\end{equation}

Here, $\Sigma_A$ and $\Sigma_B$ are the covariance matrices of $A$ and $B$ respectively, and $\Sigma_{AB}$ is the joint covariance matrix of $A$ and $B$. The mutual information can be estimated from the covariance matrices of the random variables, which can be computed from the data. This estimator of mutual information is differentiable and can be used for optimisation. We use the information measures to quantify the relationship between the reservoir states $Y_t$, the learned latent representation $V_t$, and the target output $X_{t+1}$. The learned latent representation $V_t$ is defined as the three-dimensional PCA projection of the bottleneck states $Z_t$.

\subsubsection*{Causal Emergence}
To quantify the causal emergence of learned latent representations, we use the $\Psi$ measure of causal emergence, defined by Rosas et al.~\cite{rosas2020reconciling}, which is scalable to large systems. The $\Psi$ measure quantifies the degree to which a supervenient feature $V_t$ of a system $Y_t$ is more predictive of its own future state than the sum of the information that each individual component $Y^i_t$ of the system carries about that future state $V_{t+1}$. The $\Psi$ measure is defined as,

\begin{equation}
    \Psi(V_t, Y_t) = I(V_t; V_{t+1}) - \sum_{i=1}^{N} I(Y^i_t; V_{t+1})
\end{equation}

Here, $Y^i_t$ is the state of the $i$-th reservoir neuron, and we use the learned latent representation $V_t$ as the supervenient feature of the reservoir states $Y_t$, and compute the $\Psi$ measure at each epoch of training to track how it changes during learning. While a positive value of $\Psi$ is a sufficient condition for causal emergence, in systems with a large number of components, the $\Psi$ measure can be negative. However, the $\Psi$ measure can still be used to compare the degree of emergence between different systems, and a more positive value of $\Psi$ indicates a higher degree of emergence. To track how $\Psi$ changes during training and across different trained models, we normalise it by the absolute value of the $\Psi$ measure at the first epoch of training.

\subsubsection*{Information Bottleneck}
To study the relationship between compression, emergence and generalisation, we use the information bottleneck loss to quantify the amount of predictive information about the target output $X_{t+1}$ that is retained in the learned latent representation $V_t$, while compressing away task-irrelevant information about the reservoir states $Y_t$. The information bottleneck loss is defined as,
\begin{equation}
    \mathcal{L}_{IB} = I(Y_t; V_t) - \beta I(V_t; X_{t+1}) 
\end{equation}
In this study, we use $\beta = 1$, which corresponds to an equal trade-off between compression and prediction. We compute the information bottleneck loss at each epoch of training to track how it changes during learning.

\subsection*{Representation similarity}
We compare the similarity between the learned latent representation $V_t$ and the true latent states of the dynamical system $x_t$ using Canonical Correlation Analysis (CCA). CCA finds linear combinations of the two sets of variables that are maximally correlated, and provides a measure of similarity between the two sets of variables. We compute the CCA similarity between $V_t$ and $x_t$ for both training and held-out attractors for models trained with different sizes of the bottleneck layer, and track how the similarity changes with the size of the bottleneck layer.

\subsection*{Spiking neural data analysis}
We use a previously reported~\cite{shin2019dynamics} dataset of spiking neural activity recorded using multi-tetrodes implanted in the dorsal CA1 and medial PFC of rats during a W-maze spatial learning task. All 8 rats learned the same continuous alternation rule, receiving a reward for visiting the middle arm and then the top arm after starting from the bottom arm. The rats were trained across 8 sessions, and the behavioural measures (position and velocity) and performance measures (correct and incorrect trials) were recorded using an LED at 30fps, along with the neural activity from the CA1 and the medial PFC. 

We preprocess the spiking neural data by binning the spike times into 500 ms bins and computing the firing rates of the neurons in each bin. The units with a firing rate below 0.1 Hz were excluded from the analysis. The behavioural variables were aligned to the neural data by interpolating the position and velocity of the rats using the Nadaraya-Watson kernel smoothing method. To focus on representations formed during spatial navigation, we only include timepoints when the rats were moving at a speed greater than 4 cm/s. 

The processed firing rates of the neurons from one of the regions (CA1 or medial PFC) were used as input to a simple network with a bottleneck layer and linear readout layer, and the network was trained to predict the behavioural measures (position and velocity) of the rats in the next timestep. The $\Psi$ measure of the bottleneck states was then computed separately for each session's trained model.

\subsection*{Statistical analysis}
To isolate the effect of macro predictability $I(V_t; V_{t+1})$ and the micro predictability $\sum_{i=1}^{N} I(Y^i_t; V_{t+1})$ on the next step prediction loss $\mathcal{L}$ of the network, we use a linear mixed effects model to predict $\mathcal{L}$ from the normalised macro and micro predictability, while controlling for the random effect of the different attractors, across all the trained models and both training and held-out attractors.

\begin{equation}
    \mathcal{L} = \beta_0 + \beta_1 I(V_t; V_{t+1}) + \beta_2 \sum_{i=1}^{N} I(Y^i_t; V_{t+1}) + (1|attractor)
\end{equation}

We use the MixedLM function from the statsmodels package in Python to fit the linear mixed effects model, and report the coefficients and p-values for the fixed effects.

Finally, to test the difference in the non-monotonic behaviour of the $\Psi$ measure and the MSE during learning in the spiking neural data, we fit a quadratic linear model to the measures for each subject and each brain region. We then compare the quadratic coefficients of the fitted models for the $\Psi$ measure and the MSE using a paired t-test, and report the p-values.
\begin{equation}
    M = \beta_0 + \beta_1 session + \beta_2 session^2
\end{equation}
Here $M$ is the measure of interest ($\Psi$ or MSE), and $session$ is the training session number. The magnitude of the quadratic coefficient $\beta_2$ quantifies the non-monotonicity of the measure during learning, and a larger value of $\beta_2$ indicates a more pronounced non-monotonic behaviour. The minima of the fitted model are estimated using the formula $session_{min} = -\beta_1 / (2 \beta_2)$, and the minima of the $\Psi$ measure and the MSE are compared using a permutation test, where the subject labels are permuted 10,000 times to generate a null distribution of the difference in minima between the two measures.

\bibliography{sample.bib}

\section*{Acknowledgements}

H.R. was funded by the Eric and Wendy Schmidt AI in Science Postdoctoral Fellowship, supported by Schmidt Sciences, LLC.

\appendix

\section{Training and Test Attractors}
\label{appx:attractors}
Table~\ref{tab:attractor_summary} summarises the differential equations, parameters and dynamical properties of the three training attractors (Lorenz, Rössler and Sprott B) and the three held-out test attractors (Chen, Sprott A and Lissajous) described in the Methods.

\begin{table*}[!h]
\centering
\small
\setlength{\tabcolsep}{4pt}
\renewcommand{\arraystretch}{1.25}
\begin{tabularx}{\textwidth}{
>{\raggedright\arraybackslash}p{1.9cm}
>{\raggedright\arraybackslash}X
>{\raggedright\arraybackslash}p{2.6cm}
>{\centering\arraybackslash}p{1.5cm}
>{\centering\arraybackslash}p{1.7cm}
>{\centering\arraybackslash}p{1.7cm}
>{\centering\arraybackslash}p{1.4cm}
}
\toprule
Name & Equations & Parameters & ACF decay time & Lyapunov exponent $\lambda_{\max}$ & Fractal dimension $D_{\rm KY}$& Type \\
\midrule

Lorenz &
$\begin{aligned}
\dot{x}_1 &= \sigma(x_2-x_1)\\
\dot{x}_2 &= x_1(\rho-x_3)-x_2\\
\dot{x}_3 &= x_1x_2-\beta x_3
\end{aligned}$ &
\makecell[l]{$\sigma=10$\\ $\rho=28$\\ $\beta=\tfrac{8}{3}$} &
$0.190$ & $0.906$ & $2.062$ & Training \\

R\"ossler &
$\begin{aligned}
\dot{x}_1 &= -x_2-x_3\\
\dot{x}_2 &= x_1+a x_2\\
\dot{x}_3 &= b + x_3(x_1-c)
\end{aligned}$ &
\makecell[l]{$a=0.2$\\ $b=0.2$\\ $c=5.7$} &
$0.921$ & $0.071$ & $2.013$ & Training \\

Sprott B &
$\begin{aligned}
\dot{x}_1 &= x_2 x_3\\
\dot{x}_2 &= x_1 - x_2\\
\dot{x}_3 &= 1 - x_1 x_2
\end{aligned}$ &
--- &
$1.188$ & $0.169$ & $2.144$ & Training \\

Chen &
$\begin{aligned}
\dot{x}_1 &= a(x_2-x_1)\\
\dot{x}_2 &= (c-a)x_1 - x_1x_3 + c x_2\\
\dot{x}_3 &= x_1x_2 - b x_3
\end{aligned}$ &
\makecell[l]{$a=35$\\ $b=3$\\ $c=28$} &
$0.096$ & ${\approx}2.03$ & ${\approx}2.09$ & Test \\

Sprott A &
$\begin{aligned}
\dot{x}_1 &= x_2\\
\dot{x}_2 &= -x_1 + x_2 x_3\\
\dot{x}_3 &= 1 - x_2^2
\end{aligned}$ &
--- &
$1.163$ & ${\approx}0.014$ & ${\approx}3.0$ & Test \\

Lissajous &
$\begin{aligned}
x_1(t)&=\sin(2t+\phi_1)\\
x_2(t)&=\sin(3t+\tfrac{\pi}{2}+\phi_2)\\
x_3(t)&=\sin(5t+\tfrac{\pi}{4}+\phi_3)
\end{aligned}$ &
\makecell[l]{$\omega_1=2,\;\omega_2=3$\\ $\omega_3=5$\\ $\phi_i \sim \mathcal{U}(0,2\pi)$} &
$0.348$ & $0$ & $1$ & Test \\

\bottomrule
\end{tabularx}
\caption{Dynamical systems used in this study with standard parameters. ACF decay times ($\tau^*$) are measured numerically in integration-time units for average autocorrelation to fall below $1/e$. Maximum Lyapunov exponents ($\lambda_{\max}$, nats/time) and Kaplan--Yorke fractal dimensions ($D_{\rm KY}$) are from published sources. Lissajous is periodic so $\lambda_{\max}=0$ and $D=1$ exactly.}
\label{tab:attractor_summary}
\end{table*}

\end{document}